\documentclass[11pt]{article}
\usepackage{amssymb}
\usepackage{amsmath}
\usepackage{amscd}
\usepackage{latexsym}

\catcode`@=11
\renewcommand{\section}{\@startsection{section}{1}{0pt}{\medskipamount}
{\medskipamount}{\large\bf}}
\numberwithin{equation}{section}
\catcode`@=12
\def\a{\alpha}
\def\b{\beta}

\def\de{\delta}
\def\eps{\epsilon}
\def\ve{\varepsilon}

\def\vs{\varsigma}

\newcommand{\C}{\mathbb C}
\newcommand{\R}{\mathbb R}

\newcommand{\Gcal}{{\cal G}}
\newcommand{\Acal}{{\cal A}}
\newcommand{\Ach}{{\widehat{\cal A}}}

\newcommand{\Ical}{{\cal I}}
\newcommand{\Mcal}{{\cal M}}
\newcommand{\Fcal}{{\cal F}}
\newcommand{\Fch}{{\widehat{\cal F}}}
\newcommand{\Ncal}{{\cal N}}

\newcommand{\Pcal}{{\cal P}}

\newcommand{\Ab}{{\mathbb A}}
\newcommand{\gfrak}{{\mathfrak g}}

\def\tr{\textrm{tr}}
\def\diff{\textrm{d}}
\def\pa{\mbox{$\partial$}}
\def\sfrac#1#2{{\textstyle\frac{#1}{#2}}}
\def\+{\dagger}
\def\={\ =\ }
\def\und{\qquad\textrm{and}\qquad}
\def\and{\quad\textrm{and}\quad}

\def\for{\quad\textrm{for}\quad}

\def\Id{\mathrm{Id}}

\begin{document}

\begin{titlepage}
\setcounter{page}{0}

\hspace{2.0cm}

\begin{center}

{\huge\bf
Skyrme model from 6d $\cal N$=(2,0) theory }

\vspace{12mm}

{\Large
Tatiana A. Ivanova${}^*$, \ Olaf Lechtenfeld${}^{\+\times}$ \ and \  Alexander D. Popov${}^\+$
}\\[10mm]
\noindent ${}^*${\em
Bogoliubov Laboratory of Theoretical Physics, JINR\\
141980 Dubna, Moscow Region, Russia
}\\
{Email: ita@theor.jinr.ru
}\\[5mm]
\noindent ${}^\+${\em
Institut f\"ur Theoretische Physik,
Leibniz Universit\"at Hannover \\
Appelstra\ss{}e 2, 30167 Hannover, Germany
}\\
{Email: alexander.popov@itp.uni-hannover.de}
\\[5mm]
\noindent ${}^\times${\em
Riemann Center for Geometry and Physics,
Leibniz Universit\"at Hannover \\
Appelstra\ss{}e 2, 30167 Hannover, Germany
}\\
{Email: olaf.lechtenfeld@itp.uni-hannover.de}

\vspace{20mm}

\begin{abstract}
\noindent
We consider 5d Yang--Mills theory with a compact ADE-type gauge group $G$
on $\R^{3,1}\times\Ical$, where $\Ical$ is an interval. The maximally supersymmetric extension
of this model appears after compactification on $S^1$ of 6d $\Ncal{=}\,(2,0)$ superconformal 
field theory on $\R^{3,1}\times S^2_2$, where $S^2_2\cong \Ical\times S^1$ is a two-sphere 
with two punctures. In the low-energy limit, when the length of $\Ical$ becomes small, the 5d 
Yang--Mills theory reduces to a nonlinear sigma model on $\R^{3,1}$ with the Lie group $G$
as its target space. It contains an infinite tower of interacting fields whose leading
term in the infrared is the four-derivative Skyrme term. A maximally supersymmetric generalization 
leading to a hyper-K\"ahler sigma-model target space is briefly discussed.
\end{abstract}

\end{center}
\end{titlepage}

\section {Introduction and summary}

\noindent It is generally believed that the Skyrme model~\cite{1} describes low-energy
QCD by interpreting baryons as solitons of a chiral model (see e.g.~\cite{2}
for a review and references). The standard Skyrme model describes pion
degrees of freedom, and it is not easy to include other mesons into the
model. A possible resolution of this difficulty was proposed by Sakai and
Sugimoto~\cite{3,4} who analyzed non-supersymmetric D4-D8-$\overline{\rm D8}$ brane
configurations in string theory and the holographic dual of this system.
Starting from the DBI action they arrived at gauge theory on five-dimensional
AdS-type manifold $M^5$ with Minkowski space $\R^{3,1}$ as a conformal boundary
and fifth spatial coordinate $z\in \R$ in the additional holographic direction.
Their further analysis of holographic QCD leads to an effective Skyrme model
on $\R^{3,1}$.

The group-valued Skyrme field (parametrizing the pion) in the Sakai--Sugimoto
model corresponds to the holonomy of a gauge connection,
\begin{equation}\label{1.1}
g(x) = \Pcal\exp \left(\int^\infty_{-\infty}\Acal_z(x,z) \diff z\right)\for x\in \R^{3,1} ,
\end{equation}
where the component $\Acal_z$ of a gauge potential along the holographic
direction in $M^5$ corresponds to the quark condensate of QCD, and  the
holonomy (\ref{1.1}) describes the space-time dependent fluctuations around the
manifold of vacua governed by the Skyrme model~\cite{5}. Static skyrmion fields
in this model correspond to Yang--Mills instantons on a Euclidean slice
of $M^5$.

The Sakai--Sugimoto model~\cite{3,4} is currently the best known holographic
model of hadron physics.\footnote{See e.g.~\cite{5,6} for reviews and
references.} The Skyrme Lagrangian as seen by holography is modified
-- an infinite number of terms coupling the pion field to a tower of
vector mesons naturally appears in the Lagrangian. Remarkably, the
holographic description of static baryons as instantons via holonomies of
type (\ref{1.1}) was anticipated by Atiyah and Manton~\cite{7}. 
Sutcliffe introduced a simplified version~\cite{8,9} of the Sakai--Sugimoto
model in which Yang--Mills theory is defined on 
flat $M^5 = \R^{3,1}\times \R$ but the holonomy (\ref{1.1}) is again defined along
$z\in \R$. Truncating this model one also gets the standard Skyrme model
on $\R^{3,1}$.

Here, we show that the Skyrme model (and its extension by the tower
of vector mesons) appears also from 6d $\Ncal{=}\,(2,0)$ superconformal field
theory with an ADE-type gauge group~$G$ and defined on $\R^{3,1}\times S^2_2$,
where $S^2_2$ is a two-sphere with two punctures. This theory, which 
describes multiple M5-branes, is first compactified on a circle
$S^1\hookrightarrow S^2_2\cong S^1\times \Ical$, where $\Ical$ is a closed interval,
to the five-dimensional maximally supersymmetric Yang--Mills theory.
When  $\Ical$ shrinks, it leads to a supersymmetric sigma model on $\R^{3,1}$~[10,11]. 
The target space of this sigma model is the moduli space $M^{hK}_ \Ical$
of Nahm's equations for adjoint scalar fields $\phi_A$, $A = 1,2,3$,
defined on the interval $\Ical = [-\pi, \pi]$. This $M^{hK}_ \Ical$ is also the moduli space
of vacua of the 5d SYM theory, with flat connection and other scalar fields
vanishing. In the case of Dirichlet boundary conditions this moduli space $M^{hK}_ \Ical$ 
coincides with the cotangent bundle $T^*G_\C$ carrying a hyper-K\"ahler metric~\cite{Bie, 12,10}. 
Here, $G_\C$ is a complexification of the gauge group~$G$. 
Topologically $T^*G_\C \cong G\times \gfrak\times \gfrak\times \gfrak$, where $G$ describes
the moduli space $M_\Ical$ of the gauge potential $\Acal_\Ical$ along the interval $\Ical$, and
the triple product of $\gfrak=\textrm{Lie}\,G$ describes the moduli space of the three adjoint
scalar fields~$\phi_A$ of 5d SYM theory~\cite{10,11}.

Since the action of the 6d superconformal field theory is not available,
we begin with pure Yang--Mills in five dimensions and show how the Skyrme model
appears in a low-energy limit under rather natural assumptions. 
In fact, the Skyrme term is the leading piece in a systematic expansion
of the vector meson tower contributions.
Our derivation is based on the adiabatic approach~\cite{13}--\cite{19},
which provides the expansion parameter
and differs from the holographic approach used in~\cite{3,4,8,9}. 
Finally we briefly discuss a generalization of
our results to the supersymmetric case with $\phi_A \not= 0$. 
There are many difficulties on the way to a supersymmetrization of the
Skyrme model (see e.g.~\cite{20} and references therein). 
Our approach can give a clue to the construction of an $\Ncal{=}\,2$ 
supersymmetric Skyrme model in four dimensions, which seems yet unknown.
To summarize, we demonstrate that the extended Skyrme model (describing
the pion plus the tower of vector mesons) emerges not only from a D-brane system 
of string theory but also from an M5-brane system of M-theory.

\bigskip

\section {Action functional in five dimensions}

\noindent {\bf Moduli space.} Let $M^d$ be an oriented smooth manifold of dimension $d$, $G$ a
compact ADE-type Lie group with $\gfrak$ as its Lie algebra,
$P$ a principal  $G$-bundle over   $M^d$, $\Acal$ a connection one-form on $P$
and $\Fcal =\diff \Acal + \Acal\wedge\Acal$ its curvature. We denote
by $\Ab$ the space of irreducible connections on $P$, by $\Gcal$ the infinite-dimensional
group of gauge transformations acting on $\Ab$ with the infinitesimal action
of $\Gcal$ defined by its Lie  algebra Lie$\,\Gcal$,
\begin{equation}\label{2.1}
\Gcal \ni f:\ \Acal\ \mapsto\ \Acal^{f}= f^{-1} \Acal  f + f^{-1}\diff f \und
\quad {\rm Lie}\,\Gcal \ni\eps :\  \de_\eps\Acal = D_\Acal\eps \  ,
\end{equation}
where $D_\Acal\eps :=\diff\eps + [\Acal , \eps ]$. The {\it moduli space} of connections on $P$ is defined as the quotient $\Ab /\Gcal$, i.e.~as the space of orbits of $\Gcal$
in $\Ab$.

\medskip

\noindent {\bf Space $\R^{3,1}\times\Ical$.} Now we consider $d{=}5$ and Yang--Mills theory on the direct product manifold $M^5=\R^{3,1}\times\Ical$ for $\Ical=[-\pi,\pi]$, 
with coordinates $(x^{\mu})=(x^a, x^4)$, where $x^a\in \R^{3,1}$ and $x^4\in\Ical$. 
We introduce a family of flat metrics,
\begin{equation}\label{2.2}
 \diff s^2_R \= g_{\mu\nu}^R\,\diff x^\mu \diff x^\nu \= \eta_{ab}\,\diff x^a \diff x^b + R^2(\diff x^4)^2\ ,
\end{equation}
where $(\eta_{ab})={\rm diag} (-1,1,1,1)$ with $a,b=0,1,2,3$, and the dimensionful coordinate $\tilde x^4=R x^4$ parametrizes the scaled interval
$\Ical_R=[-\pi R, \pi R]$ of length $2\pi R$.

\medskip

\noindent {\bf Gauge fields.} Let us look at the principal $G$-bundle $P$ over $\R^{3,1}\times\Ical$
with  a gauge potential (connection) $\Acal$ and the gauge field  (curvature) $\Fcal$ both valued in the Lie algebra $\gfrak$ of the
group $G$. On  $\R^{3,1}\times\Ical$ we have the obvious splitting
\begin{equation}\label{2.3}
\Acal \= \Acal_{a}\,\diff x^a+\Acal_{4}\,\diff x^4\und
\Fcal \=\sfrac12\Fcal_{ab}\,\diff x^a \wedge \diff x^b + \Fcal_{a4}\,\diff x^a \wedge \diff x^4\ .
\end{equation}
For the generators $I_i$ in the adjoint representation of $G$ we will use the standard normalization $\tr (I_iI_j) = -2\de_{ij}$ with
$i,j=1,\ldots,{\rm dim}G$.

For the metric tensor (\ref{2.2}) we have $(g^{\mu\nu}_R)=(\eta^{ab}, R^{-2})$ and
$\det(g_{\mu\nu}^R)=-R^{2}$. We denote by $\Fcal^{\mu\nu}_R$
the contravariant components raised from $\Fcal_{\mu\nu}$ by the tensor $g^{\mu\nu}_R$ and by
$\Fcal^{\mu\nu}$ those obtained by $g^{\mu\nu}\equiv g^{\mu\nu}_{R=1}$.
We have $\Fcal^{ab}_R=\Fcal^{ab}$ and $\Fcal^{a4}_R=R^{-2}\Fcal^{a4}$.

\medskip

\noindent {\bf Action.}
The standard Yang--Mills action functional takes the form
\begin{equation}\label{2.4}
S\=-\frac{1}{8e^2}\int_{\R^{3,1}\times \Ical} \!\!\!\!\!\!\diff^5x\ \sqrt{|\det g^R|}\,\tr\Fcal_{\mu\nu}\Fcal^{\mu\nu}_R
\= -\frac{1}{8e^2R}\int_{\R^{3,1}\times \Ical} \!\!\!\!\!\!\diff^5x\ \tr\bigl( R^2\,\Fcal_{ab}\Fcal^{ab}+2\Fcal_{a4}\Fcal^{a4}\bigr)\ ,
\end{equation}
where $e$ is the gauge coupling constant.
A supersymmetric extension of (\ref{2.4}) can be found e.g.~in \cite{11}. 
We want to discuss the infrared region of the pure Yang--Mills
model~(\ref{2.4}), because QCD is not a supersymmetric theory and we are after the Skyrme model as a description of hadrons.
The infrared is reached by tuning down the parameter~$R$, so the interval $\Ical_R$ becomes very short.

\bigskip

\section{Moduli space of vacuua}

\noindent {\bf Gauge group.} Consider the group $\Gcal = C^\infty (\R^{3,1}\!\times\Ical,G)$
and its restriction $\Gcal_{\Ical}$ to ${\Ical}$ by fixing $x^a\in \R^{3,1}$ to an arbitrary value.
The boundary of our manifold $M^5=\R^{3,1}\times\Ical$ consists of two Minkowski spaces at
$x^4=\pm\pi$. On
manifolds $M^d$ with nonempty boundary $\pa M^d$, the group of gauge transformations is
naturally restricted
to the identity when $x^a$ reaches $\pa M^d$ (see e.g.~\cite{21} and \cite{10}-\cite{12})
for our case).
For our $M^5$, this means allowing only gauge-group elements $f\in \Gcal$ obeying
$f(\pa M^5)=\Id$ on $\pa M^5=\R^{3,1}_{x^4=\pm\pi}$.
We denote this group by $\Gcal^0$ and its restriction to $\Ical$ by $\Gcal^0_\Ical$.

\medskip

\noindent {\bf Vacua.} Vacua of Yang--Mills theory  (\ref{2.4}) on $M^5=\R^{3,1}\times\Ical$ are
defined by the vanishing of the gauge fields, $\Fcal =0$. 
The components $\Fcal_{ab} =0$ can be solved by putting $\Acal_{a} =0$, and from $\Fcal_{a4} =0$ one obtains
\begin{equation}\label{3.1}
\Acal_z\=\Acal_z(z)\=h^{-1}\pa_z h\ ,
\qquad\textrm{with}\quad z=x^4 \and \Acal_{z}=\Acal_4
\end{equation}
for notational convenience. 
Here $h(z)\in\Gcal_\Ical$ is not an element of the gauge group $\Gcal^0_\Ical$.
Therefore, $\Acal_{z}$ in~(\ref{3.1}) cannot be transformed to zero by an admissible gauge transformation.
In fact, 
\begin{equation}
h(-\pi )=: h_L\in G_L=\Gcal_\Ical\big|_{z=-\pi}\cong G \und
h(\pi )=: h_R\in G_R=\Gcal_\Ical\big|_{z=\pi}\cong G \ .
\end{equation}

\medskip

\noindent {\bf Holonomy.} For the interval $\Ical =[-\pi , \pi ]$ with coordinate $z$ we
denote by $\Acal_{\Ical}=\Acal_z\diff z$
a connection one-form on the bundle $P_\Ical =\Ical\times G\to\Ical$ over $\Ical$, which is a
restriction of the bundle $P=\R^{3,1}\times\Ical\times G\to\R^{3,1}\times\Ical$ to $\Ical$
by fixing an arbitrary point $x^a\in \R^{3,1}$. 
Then, given any connection  $\Acal_{\Ical}$ on $\Pcal_{\Ical}$
we have the differential equation  (\ref{3.1}) for $h$. 
The group of gauge transformations $\Gcal^0_\Ical$ acts on $\Acal_{\Ical}$ and $h$ by
\begin{equation}\label{3.2}
\Gcal^0_\Ical\ni f :\  \Acal^{}_{\Ical}\ \mapsto\ 
\Acal^{f}_{\Ical}= f^{-1} \Acal^{}_{\Ical} f + f^{-1}\diff_{\Ical} f \und
h\ \mapsto\ h^f=f^{-1}(\pi )h(z)f(z)=hf\ ,
\end{equation}
with $\diff_{\Ical}=\diff z\pa_z$ and $f(\pi)=\Id$ for $f\in\Gcal^0_\Ical$.

To solve (\ref{3.1}) with Dirichlet boundary conditions~\cite{10, 11} 
one has to choose an initial value for the $G$-valued function $h$ on $\Ical$.
From (\ref{3.1}) it follows that $h$ and $h_{L}^{-1}h$ define the same connection $\Acal_z\diff z$,
since $h_{L}$ does not depend on $z\in\Ical$.
Hence, the space of all flat connections $\Acal^{}_{\Ical}$ on $P^{}_{\Ical}$
(equivalently, the space of all solutions $h$ to  (\ref{3.1}) with fixed initial condition)
is the coset space $\Ncal_{\Ical}=\Gcal_\Ical /G_L$. The unique solution to the differential equation
(\ref{3.1}) can be written as
\begin{equation}\label{3.4}
h(z)\=\Pcal \exp\Bigl(\int^z_{-\pi} A_y\,\diff y\Bigr)\ ,
\end{equation}
where $\Pcal$ denotes path ordering. Notice that for  (\ref{3.4}) we have $h(z{=}{-}\pi)=\Id$, 
i.e.~$h\in \Ncal_{\Ical}$. The group element
$g{=}h(z{=}\pi )\in G_R\cong G$
is the holonomy of $\Acal_{\Ical}$, which is not transformed under the group  $\Gcal^0_{\Ical}$ of
gauge transformations, as follows from  (\ref{3.2}) and $f(\pi)=\textrm{Id}$  for $f\in \Gcal^0_\Ical$.

\medskip

\noindent {\bf Gauge-equivalent vacua.}  We note that
$\Gcal_{\Ical} =\Gcal^0_{\Ical}\rtimes (G_L\times G_R)$,
and the solution space of (\ref{3.1})
is $\Ncal_{\Ical}=\Gcal^0_{\Ical}\rtimes G$ from $G\cong G_R$. 
Thus the gauge group $\Gcal^0_{\Ical}$
can be defined as the kernel of the projection (evaluation map)
\begin{equation}\label{3.6}
q:\ \Ncal_{\Ical} \stackrel{\Gcal^0_\Ical}{\to} G
\qquad\textrm{with}\qquad
h(z)\mapsto h(\pi )\ .
\end{equation}
The action (\ref{3.2}) of $\Gcal^0_{\Ical}$ on $\Ncal_\Ical$ is free, and the projection $q$ is injective. 
Hence, (\ref{3.6}) is the principal  $\Gcal^0_{\Ical}$-bundle over $G$. 
The base $G$ of this bundle is the moduli space $\Mcal_{\Ical}$ of vacua of Yang--Mills
theory on $\R^{3,1}\times\Ical$.

\bigskip

\section{Changes of $\Acal_\Ical$ under shifts on $\Mcal_{\Ical}$}

\noindent {\bf Connections on $G$.} 
Introducing local coordinates $X=\{X^\a\}$ on $G$, the differential of the holonomy element $g=h(\pi)\in G$
can be expressed as $\diff g=(\partial_\alpha g)\diff X^\alpha$.
Then, the canonical flat connection on the tangent bundle~$TG$ reads
\begin{equation}\label{4.1}
\Gamma \=g^{-1}\diff g\=(g^{-1}\pa_\a g)\,\diff X^\a \=e^i_\a I_i\,\diff X^\a \= e^iI_i\ ,
\end{equation}
where $e^i$ are left-invariant one-forms on $G$. They satisfy the Maurer--Cartan equations
\begin{equation}\label{4.2}
\diff e^i + \sfrac12\, f^i_{jk} e^j\wedge e^k =0\ ,
\end{equation}
where $ f^i_{jk}$ are the structure constants of the group~$G$. The collection $\{e^i\}$ forms an
orthonormal basis on the cotangent bundle $T^*G$, and for
a metric on $G$ we have
\begin{equation}\label{4.3}
\diff s^2_G\=\de_{ij}e^ie^j \= \de_{ij}e^i_\a e^j_\b\, \diff X^\a \diff X^\b
\ =:\ g_{\a\b}\,\diff X^\a\diff X^\b \ .
\end{equation}

\medskip

\noindent {\bf Variation of $\Acal_{\Ical}$}. 
Our fields $g$, $h$ and $\Acal_{\Ical}$ are parametrized by the coordinates $X^{\a}$ of~$G$.
In general, $\Acal_{\Ical}$ belongs to the space $\Ncal_{\Ical}$ described in Section~3 
and fibred over~$G$. 
We introduce the tangent bundle $T\Ncal_{\Ical}$ of $\Ncal_{\Ical}$ as the fibration
\begin{equation}\label{4.6}
q_*: \ T\Ncal_{\Ical}\ \to\ TG
\end{equation}
with fibres  $T^{}_{\Acal_{\Ical}}\Gcal^0_\Ical\cong\,$Lie$\,\Gcal^0_\Ical$ at any point $\Acal_{\Ical}\in G$.
Also, we have $T^{}_{\Acal_{\Ical}}G\cong\gfrak$ and therefore
\begin{equation}\label{4.7}
T^{}_{\Acal^{}_{\Ical}} \Ncal_{\Ical}\=q^*  T^{}_{\Acal_{\Ical}} G\oplus T^{}_{\Acal_{\Ical}}\Gcal^0_\Ical
\ \cong\ \gfrak\oplus {\rm Lie}\,\Gcal^0_\Ical\ .
\end{equation}
Note that even if $\Acal_{\Ical}$ belongs to the base $G$ of the fibration (\ref{3.6}) (after fixing a gauge),
its derivative $\pa_\a\Acal_{\Ical}$ with respect to $X^\a$ belongs to the tangent space $T^{}_{\Acal_{\Ical}}\Ncal_{\Ical}$
and not necessarily to the tangent space $T^{}_{\Acal_{\Ical}} G$. However, $\pa_\a\Acal_{\Ical}$ can always be decomposed as
\begin{equation}\label{4.8}
\pa_\a\Acal_{z}\=\de_{\a}\Acal_{z}+\de^{}_{\eps_\a}\Acal_{z}\ =:\ \xi_\a + D_z\eps_\a
\qquad\textrm{with}\quad \xi_\a\in T^{}_{\Acal_{\Ical}}G \and D_z\eps_\a\in T^{}_{\Acal_{\Ical}}\Gcal^0_\Ical\ ,
\end{equation}
where $T^{}_{\Acal_{\Ical}}G\cong\gfrak$ and $T^{}_{\Acal_{\Ical}}\Gcal^0_\Ical\cong{\rm Lie}\,\Gcal^0_\Ical$. 
The $\gfrak$-valued gauge parameters $\eps_\a$ generate infinitesimal gauge transformations which, 
after the $\pa_\a$-shift, bring $\Acal_{\Ical}$ back to $G$.

Orthogonality of $\xi_\a=\de_{\a}\Acal_{z}$ and $D_z\eps_\a=\de^{}_{\eps_\a}\Acal_{z}$ is achieved by imposing the
condition
\begin{equation}\label{4.9}
D_z\xi_{\a}=0\quad\Leftrightarrow\quad D^2_z\eps_\a=D_z\pa_\a\Acal_z\ .
\end{equation}
From (\ref{3.1}) and (\ref{4.9}) one obtains
\begin{equation}\label{4.10}
\xi_{\a}\=e^i_\a h^{-1}I_i\,h \=  h^{-1}(g^{-1} \pa_\a g) h\ ,
\end{equation}
which shows that the $z$-dependence of $\xi_{\a}$ is located in $h(z)$ alone.

\bigskip

\section{Skyrme model in the infrared limit of 5d Yang--Mills}

\noindent {\bf Moduli-space approximation.} After having described the moduli space $\Mcal_{\Ical}$ of
Yang--Mills theory on $\R^{3,1}\times\Ical$, we return to non-vacuum gauge fields.
In the moduli-space approximation it is postulated that the collective coordinates $X^\a$ depend
on $x^a\in \R^{3,1}$, so that $X^\a = X^\a (x^a)$ may be considered as dynamical fields,
and that this captures the $x^a$ dependence of ``slow'' full solutions.
The low-energy effective action for $X^\a$ is derived by expanding
\begin{equation}\label{5.1}
\Acal_{\mu} \= \Acal_{\mu}(X^\a (x^a), x^4) + \ldots\ ,
\end{equation}
where the first term depends on $x^a\in \R^{3,1}$ only via the coordinates $X^\a\in\Mcal_{\Ical}$~\cite{13,14,16,19}.
Then for distances in $\R^{3,1}$ which are large in comparison with the length $2\pi R$ of the interval $\Ical_R$ 
(or, in other words, for small values of $R$) all terms in~(\ref{5.1}) beyond the first one are discarded. 
By substituting the leading term of~(\ref{5.1}) into the initial action~(\ref{2.4}), one obtains
an effective field theory describing small fluctuations around the vacuum manifold.

\medskip

\noindent {\bf Kinetic part of effective action}. 
The gauge potential decomposes as
\begin{equation}
\Acal = \Acal^{}_{\R^{3,1}} + \Acal_{\Ical} \qquad\textrm{with}\quad
\Acal^{}_{\R^{3,1}} = \Acal_a \diff x^a \and \Acal_{\Ical} = \Acal_z \diff z\ .
\end{equation}
For any fixed $x^a\in \R^{3,1}$,
the part  $\Acal_{\Ical}(X^\a(x^a), x^4)$ belongs to the space $\Ncal_{\Ical}$ described
in Sections 3 and~4. We now use the formul\ae\ from these sections and include the
dependence on $x^a$. In particular, multiplying (\ref{4.8}) by $\pa_a X^\a$, we obtain
\begin{equation}\label{5.2}
\pa_a \Acal_{z} \= (\pa_a X^\a)\xi_{\a}+D_z\eps_a\ ,
\end{equation}
where $\eps_a= (\pa_a X^\a)\eps_{\a}$  is the pull-back of $\eps_\a$ to $\R^{3,1}$.

We have provided the details of the $\Acal_{\Ical}$ part of the connection~$\Acal$.
On the other hand, the components $\Acal_a$ of the $\Acal^{}_{\R^{3,1}}$ part are not yet fixed. 
To this end, we note that
\begin{equation}\label{5.3}
\Fcal_{a4}\=\pa_a\Acal_z - D_z\Acal_a\= (\pa_a X^\a)\xi_{\a} +D_z(\eps_a{-}\Acal_a)\ .
\end{equation}
In the moduli-space approximation, $\Fcal_{a4}$ has to be tangent to $\Mcal_{\Ical}$ (see e.g.~\cite{13,14}).
Hence, the second term in (\ref{5.3}) should vanish, i.e.~$\eps_a{-}\Acal_b$ must lie in kernel of $D_z$,
which according to (\ref{4.9}) is proportional to $\xi_\a$. Thus, we have
\begin{equation}\label{5.4}
\Acal_a \= \eps_a + A_a^\a\xi_\a \= \eps_a + A^i_a(X)h^{-1}I_i\,h \ ,
\end{equation}
where $A^i_a$ are arbitrary functions of the group coordinates $X^\a =X^\a (x^a)$.
For simplicity we pick a gauge where $A^i_a=0$, so that 
\begin{equation}\label{5.4a}
\Acal_a \= \eps_a \qquad\textrm{with boundary conditions}\quad
\eps_a(z{=}{-}\pi) = 0 = \eps_a(z{=}\pi)\ .
\end{equation}

Substituting
\begin{equation}\label{5.5}
\Fcal_{a4}\=(\pa_a X^\a)\xi_{\a} \= (\pa_a X^\a)\,e^i_\a h^{-1}I_i\,h \= h^{-1}(g^{-1}\pa_a g) h
\end{equation}
into the action~(\ref{2.4}), the second term becomes
\begin{equation}\label{5.6}
S_{\textrm{kin}} \=
-\frac{1}{8 e^2 R}\int_{\R^{3,1}\times \Ical} \!\!\!\!\diff^5x\ \eta^{ab}\ \tr\Fcal_{a4}\Fcal_{b4}\=
-\frac{\pi}{4e^2 R}\int_{\R^{3,1}} \!\diff^4x\ \eta^{ab}\, \tr (L_a L_b)\ ,
\end{equation}
where we used (\ref{5.5}) and the definition
\begin{equation}
L_a\ :=\ g^{-1}\pa_a g\ .
\end{equation}
Thus, this part of the action reduces to a sigma model on $\R^{3,1}$ with $\Mcal_{\Ical}\cong G$ as target space.

\medskip

\noindent {\bf Skyrme term}. 
For calculating the first term in the action (\ref{2.4}) it is convenient to rewrite (\ref{5.4a}) as
\begin{equation}\label{5.7}
\Acal_a \= \eps_a \= h\,(h^{-1}\eps_a h+h^{-1}\partial_a h)\,h^{-1} + h\,\partial_a h^{-1}
\ =:\ h\,\Ach_a h^{-1} + h\,\partial_a h^{-1}\ ,
\end{equation}
where $\Ach_a$ depends on $z$.
The boundary conditions (\ref{5.4a}) for $\eps_a$ translate to
\begin{equation}
\Ach_a(z{=}{-}\pi) = 0 \und \Ach_a(z{=}\pi) = L_a
\end{equation}
since $h(z{=}{-}\pi) =\textrm{Id}$ and $h(z{=}\pi)=g$.
Therefore, we can expand $\Ach_a(z)$ on $\Ical$ as\footnote{
The coefficient linear in $z$ is just a convenient choice of a function 
interpolating between $0$ and $1$ on $\Ical$.
It yields a family of metric-compatible linear connections which are non-flat inside $\Ical$,
with torsion $T^i_{jk}=\frac{z}{\pi}f^i_{jk}$ 
and curvature $R_{ijkl} = \frac{z^2-\pi^2}{2\,\pi^2}\,\de_{mr}\, f^m_{ij}\, f^r_{kl}$
(see e.g.~\cite{23} for a discussion). At $z{=}0$ one finds the Levi-Civita connection.}
\begin{equation}
\Ach_a(z) \= \sfrac{z+\pi}{2\,\pi}\,L_a + \sum_{n=1}^\infty B_a^{(n)} \sin nz\ ,
\end{equation}
where $L_a$ represents the pion degree of freedom 
and the $B_a^{(n)}$ describe the tower of mesons.

The curvature of $\Ach$ then computes to
\begin{equation}
h^{-1}\Fcal_{ab}\,h \= \Fch_{ab} \= \partial_a\Ach_b-\partial_b\Ach_a+[\Ach_a,\Ach_b]
\= \sfrac{z^2-\pi^2}{4\,\pi^2}\,[L_a,L_b] + B_{ab}\ ,
\end{equation}
where the term $B_{ab}$ contains the meson contributions.
Substituting this into the action~(\ref{2.4}) and truncating to the pion, 
i.e.~discarding all $B_{ab}$ terms, we obtain
\begin{equation}\label{5.11}
S_{\textrm{Skyrme}} \= 
-\frac{R}{8 e^2}\int_{\R^{3,1}\times \Ical} \!\!\!\!\diff^5x\ \tr\Fcal_{ab}\Fcal^{ab} \=
-\frac{\pi\,R}{120 e^2} \int_{\R^{3,1}} \!\diff^4x\ \eta^{ac}\eta^{bd}\,\tr\bigl([L_a, L_b][L_c, L_d]\bigr) \ .
\end{equation}

Thus, in the infrared limit the Yang--Mills action on $\R^{3,1}\times \Ical$ is reduced to the effective action
of the Skyrme model,
\begin{equation}\label{5.12}
S_{\textrm{eff}} = -\int_{\R^{3,1}}\!\!\diff^4x\biggl\{\frac{f^2_\pi}{4}\,\eta^{ab}\, \tr (L_a L_b) +
\frac{1}{32\vs^2}\,\eta^{ac}\eta^{bd}\,\tr\bigl([L_a, L_b][L_c, L_d]\bigr)\biggr\}\ ,
\end{equation}
where $\vs$ is the dimensionless Skyrme parameter and  $f_{\pi}$ may be interpreted as
the pion decay constant.
Their relation to the gauge coupling and the infrared scale~$R$ is
\begin{equation}\label{5.13}
\frac{f_\pi^2}{4} = \frac{\pi}{4 e^2 R}  \und  \frac{1}{32\vs^2} = \frac{\pi R}{120 e^2}\ .
\end{equation}
We see that the ratio of these parameters depends on the length of the interval $\Ical_R=[-\pi R,\pi R]$
characterizing the approach to the infrared.

\medskip

\noindent {\bf Towards to supersymmetric model}. What will change if we consider the infrared limit of
maximally supersymmetric Yang--Mills theory (SYM)? 5d SYM contains five adjoint scalars, namely $\phi^A$, $\phi^4$,
$\phi^5$, and the moduli space $\Mcal^{hK}_\Ical$ of this theory is defined as the moduli
space of flat connections $\Fcal_{\mu\nu}=0$, which we considered, extended by the moduli space of solutions
to the Nahm equations
\begin{equation}\label{5.14}
\pa_z\phi^A + [\Acal_z, \phi^A]\=\sfrac12\ve^A_{BC}[\phi^B, \phi^C]\und \phi^4=\phi^5=0\ ,
\end{equation}
on the scalar fields depending on $z\in\Ical=[-\pi,\pi]$ with all fermions vanishing~\cite{10,11}.
This moduli space $\Mcal^{hK}_\Ical$ depends essentially on the boundary conditions imposed on $\Acal_\Ical$
and $\phi^A$ and was discussed e.g.~in~\cite{Bie, 12} (see also references therein). For the simplest Dirichlet
boundary conditions~\cite{10, Bie, 12} and assuming regularity at $z=\pm\pi$, the moduli space  $\Mcal^{hK}_\Ical$
is the cotangent bundle $T^*G_\C\cong G_\C\times \gfrak_\C\cong G\times\gfrak\times\gfrak\times\gfrak$
with a hyper-K\"ahler metric. The explicit form of the $\Ncal{=}\,2$ supersymmetric sigma-model action for the
hyper-K\"ahler vacuum moduli space $\Mcal^{hK}_\Ical$ e.g.~in~\cite{11}. Its
derivation from 5d SYM in the infrared limit is similar to the one for the bosonic case.

Finally, the Skyrme term should get supersymmetrized. We had $X^\a\in\Mcal_\Ical\cong G$. 
In terms of $X^\a$ and one-forms $e^i=e^i_\a\diff X^\a$ on $G$ the standard Skyrme term in the Lagrangian
of (\ref{5.11})  can be  written as
\begin{equation}\label{5.15}
\eta^{ac}\eta^{bd}\,\pa_aX^\a \pa_bX^\b \pa_cX^\gamma \pa_dX^\de\,  e^i_\a e^j_\b e^k_\gamma e^l_\de\,
R_{ijkl}\=\pa_aX^\a \pa_bX^\b\, \pa^a X^\gamma \pa^b X^\de\, R_{\a\b\gamma\de}\ ,
\end{equation}
where $\pa^a=\eta^{ac}\pa_c$ and $R_{ijkl}$ is the curvature of the connection on
$\Mcal_\Ical\cong G$. We expect that for the supersymmetric case (\ref{5.15}) will have the same form
but with $(X^\a,e^i,R_{ijkl})$ being defined on the hyper-K\"ahler moduli space $\Mcal^{hK}_\Ical$.
Additional fermionic and possibly auxiliary terms may also need to be deduced. However, this
is beyond the scope of our paper.

\bigskip

\noindent {\bf Acknowledgements}

\noindent
This work was partially supported by the Deutsche Forschungsgemeinschaft grant LE 838/13
and by the Heisenberg-Landau program.
It is based upon work from COST Action MP1405 QSPACE, supported by COST (European Cooperation
in Science and Technology).

\end{document}